\documentclass[a4paper,11pt]{article}
\usepackage{geometry}
\usepackage[svgnames]{xcolor}
\usepackage[colorlinks=true,urlcolor=black,linkcolor=blue,citecolor=blue,bookmarks=false]{hyperref}
\usepackage{amsfonts,amsmath,amssymb}
\usepackage{graphicx}
\usepackage{dsfont}
\usepackage{graphicx}
\usepackage{bm}
\usepackage[font=small,labelfont=bf]{caption}
\renewcommand*{\thepage}{\footnotesize\arabic{page}}
\usepackage{color}
\usepackage{enumitem}
\usepackage{boldline}
\usepackage{changepage}
\usepackage{cite}
\usepackage{textcomp}
\usepackage[title]{appendix}
\usepackage{url}
\usepackage[OT1]{fontenc}
\usepackage{authblk}
\usepackage{breakurl}
\usepackage{fancyhdr}

\title{\bf Quasi-stationary states in temporal correlations for traffic systems: Cologne orbital motorway as an example }

\author{Shanshan Wang \thanks{shanshan.wang@uni-due.de}, Sebastian Gartzke, Michael Schreckenberg and Thomas Guhr}
\affil{\textit{Fakult\"at f\"ur Physik, Universit\"at Duisburg--Essen, Lotharstra\ss e 1, 47048 Duisburg, Germany}}

\date{\today}
 
\begin{document}
\maketitle

\noindent {\bf Abstract.}
Traffic systems are complex systems that exhibit non-stationary characteristics. Therefore, the identification of temporary traffic states is significant. In contrast to the usual correlations of time series, here we study those of position series, revealing structures in time, i.e. the rich non-Markovian features of traffic. Considering the traffic system of the Cologne orbital motorway as a whole, we identify five quasi-stationary states by clustering reduced-rank correlation matrices of flows using the $k$-means method. The five quasi-stationary states with nontrivial features include one holiday state, three workday states and one mixed state of holidays and workdays. In particular, the workday states and the mixed state exhibit strongly correlated time groups shown as diagonal blocks in the correlation matrices. We map the five states onto reduced-rank correlation matrices of velocities and onto traffic states where free or congested states are revealed in both space and time. Our study opens a new perspective for studying traffic systems. This contribution is meant to provide a proof of concept and a basis for further study. 
\vspace{0.5cm}

\vspace{1cm}

\noindent\rule{\textwidth}{1pt}
\vspace*{-1cm}
{\setlength{\parskip}{0pt plus 1pt} \tableofcontents}
\noindent\rule{\textwidth}{1pt}

\section{Introduction}
\label{sec1}

A complex system~\cite{Ladyman2013,Insight2001} is a system that contains many interacting constituents or agents and is usually non-stationary, i.e., far away from any form of equilibrium. The collective behavior of individuals rather than a collection of individual behaviors is an important feature, since the complexity of the system originates from the interactions, interdependency and relationships among components~\cite{Bar2002}. One example of a complex system is a financial market, in which stocks can be understood as constituents. Only when the financial market is considered as a whole system can the characteristic collective behavior be found~\cite{Gopikrishnan2001,Plerou2002,Wang2016a,Wang2016b,Benzaquen2017 }. Other examples of complex systems are the human brain~\cite{Telesford2011}, power systems~\cite{Messina2009}, ecosystems~\cite{Levin1998}, the global climate~\cite{Rind1999} and so on. Here we focus on traffic systems, where the motions of vehicles are strongly influenced by their neighbours. The behavior of traffic flow, e.g., free flow and congested flow~\cite{Kerner2012}, has been studied extensively by modeling and simulation~\cite{Nagel1992,Schadschneider1993,Lovaas1994,Schreckenberg1995,Hoogendoorn2001,Wong2002,Fellendorf2010,Treiber2013}. While many theoretical studies have been devoted to traffic systems, empirical studies~\cite{Kerner2002,Bertini2005,Schonhof2007,Kerner2012} have been less frequent for various reasons and an improvement to this situation is urgently called for, in particular by studies of environmental influences, e.g. weather, seasons, road construction and holidays.

The non-stationary characteristics of traffic systems are seen in the time series, the mean and the variance of the flow, velocity and density changes over time. Non-stationary or relevant quasi-stationary characteristics in time series have been studied extensively in financial markets~\cite{Schafer2010,Munnix2012,Schafer2015,Rinn2015,Stepanov2015,Heckens2020}, power systems~\cite{Messina2009}, climates~\cite{Mann2004,Cheng2014}, speech recognition~\cite{Cohen2001,Rangachari2006}, traffic systems~\cite{Blandin2013,Cassidy1998}, etc. For financial markets, further studies based on non-stationary time series distinguish the markets at different time periods by specific market states in the correlation structure, through which the system passes, and in which the system remains for shorter or longer times~\cite{Munnix2012,Chetalova2015a,Chetalova2015b,Guhr2015,Heckens2020}. Thus, transitions between different market states can be revealed to get a better understanding of the system, resulting in a characteristic trajectory of the system in the space of these states. The better one understands a system, the more one is able to control the system or predict its behavior. For instance, with the states identified for previous time periods, the probabilities of the occurrence of states at later times can be estimated and methods such as Markov models~\cite{Howard2012} or hidden Markov models~\cite{Rabiner1986} may be applied. Such knowledge might provide means to identify the precursors of negative or positive developments. In the case of traffic systems, we wish to explore the potential to predict, plan and optimize traffic to maximize the flow and minimize the occurrence of traffic jams.

Due to the availability of a huge amount of traffic data nowadays, the use of clustering to identify such states in the correlation structure is possible, but has not yet been described in the literature. Here, we consider the Cologne orbital motorway in Germany as a whole and identify quasi-stationary states in time based on the data from inductive loop detectors. For this purpose, we calculate reduced-rank correlation matrices between traffic flows for different time steps and apply a clustering method to classify the reduced-rank correlation matrices. As a result, five quasi-stationary states are identified with specific features that we will reveal. We chose Cologne motorway as it is essentially a ring, limiting the complexity as compared to a freeway network.

This paper is organized as follows. We introduce the data set used in this study in section~\ref{sec2}. We then describe the method of spectrum filtering to obtain reduced-rank correlation matrices with specific eigenvalues in section~\ref{sec3}. Using $k$-means clustering for all the reduced-rank correlation matrices with large eigenvalues, we identify five quasi-stationary states and describe their features in section~\ref{sec4}. In section~\ref{sec5}, we map the five states onto reduced-rank correlation matrices of velocities and onto traffic states where free or congested states are shown in both time and space. We conclude with our results in section~\ref{sec6}.

\section{Datasets}
\label{sec2}

\begin{figure}[tbp]
\begin{center}
\includegraphics[height=0.45\textwidth, width=0.6\textwidth]{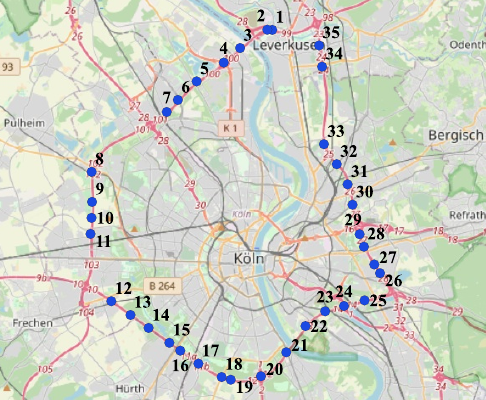}
\caption{Locations of traffic detectors at cross-sections of the Cologne orbital motorway (road numbers A1 (sections 1--11), A3 (sections 26--35), and A4 (sections 12--25)), with blue markers numbered counterclockwise. The Cologne orbital motorway we studied includes the counterclockwise direction as well. The map was developed with QGIS 3.4. The base map and data are from OpenStreetMap and OpenStreetMap Foundation. OpenStreetMap data \copyright~OpenStreetMap contributors, licensed under the Open Database License (ODbL)}
\label{fig1}
\end{center}
\end{figure}

The traffic data we used is from inductive loop detectors of 35 available cross-sections of Cologne orbital motorway (road numbers A1, A3 and A4) in Germany, as shown in figure~\ref{fig1}. Due to a lack of available data for sections in the clockwise direction in 2015, we only use the data for sections in the counterclockwise direction, which was also used for the study in~\cite{Krause2017}. The data includes the information for flows, i.e. the number of vehicles that pass a detector in a time unit, and velocities for the whole year of 2015. Since the flows and velocities are measured or averaged over one minute and then converted into vehicles/h and km/h, the data we obtained has a time resolution of one minute. For each section $k$, we combine the flows $q_{kl}(t)$ in multiple lanes $l$ into one effective lane, such that
\begin{equation}
q_k(t)=\sum_l q_{kl}(t)  \ 
\label{eq2.1}
\end{equation}
is the combined flow in section $k$. Taking account of different flows in different lanes, a simple average of the velocities over the different lanes would distort the behavior of the velocities through a whole day. To overcome this problem, we resort to the flow density of each lane
\begin{equation}
\rho_{kl}(t)=\frac{q_{kl}(t)}{v_{kl}(t)} \ ,
\label{eq2.2} 
\end{equation}
which indicates the number of vehicles in a unit distance, so as to obtain the corresponding velocities $v_{k}(t)$ by
\begin{equation}
v_k(t)=\frac{q_k(t)}{\sum_l{\rho_{kl}(t)}} \ .
\label{eq2.3}
\end{equation}
As fewer vehicles pass through some sections in some of the small time intervals, especially during the night, we aggregate the data into time intervals of 15 min by summing over all the flows and averaging over all the velocities.

\section{Correlation matrices}
\label{sec3}

In section~\ref{sec31}, we describe the correlation matrices and the method of spectrum filtering, which results in reduced-rank correlation matrices with specific eigenvalues. We explore the role of large eigenvalues by dissecting data matrices with singular value decomposition in section~\ref{sec32}.

 \subsection{Correlation matrices and their spectrum filtering}
 \label{sec31}
 
\begin{figure}[htbp]
\begin{center}
\includegraphics[width=0.8\textwidth]{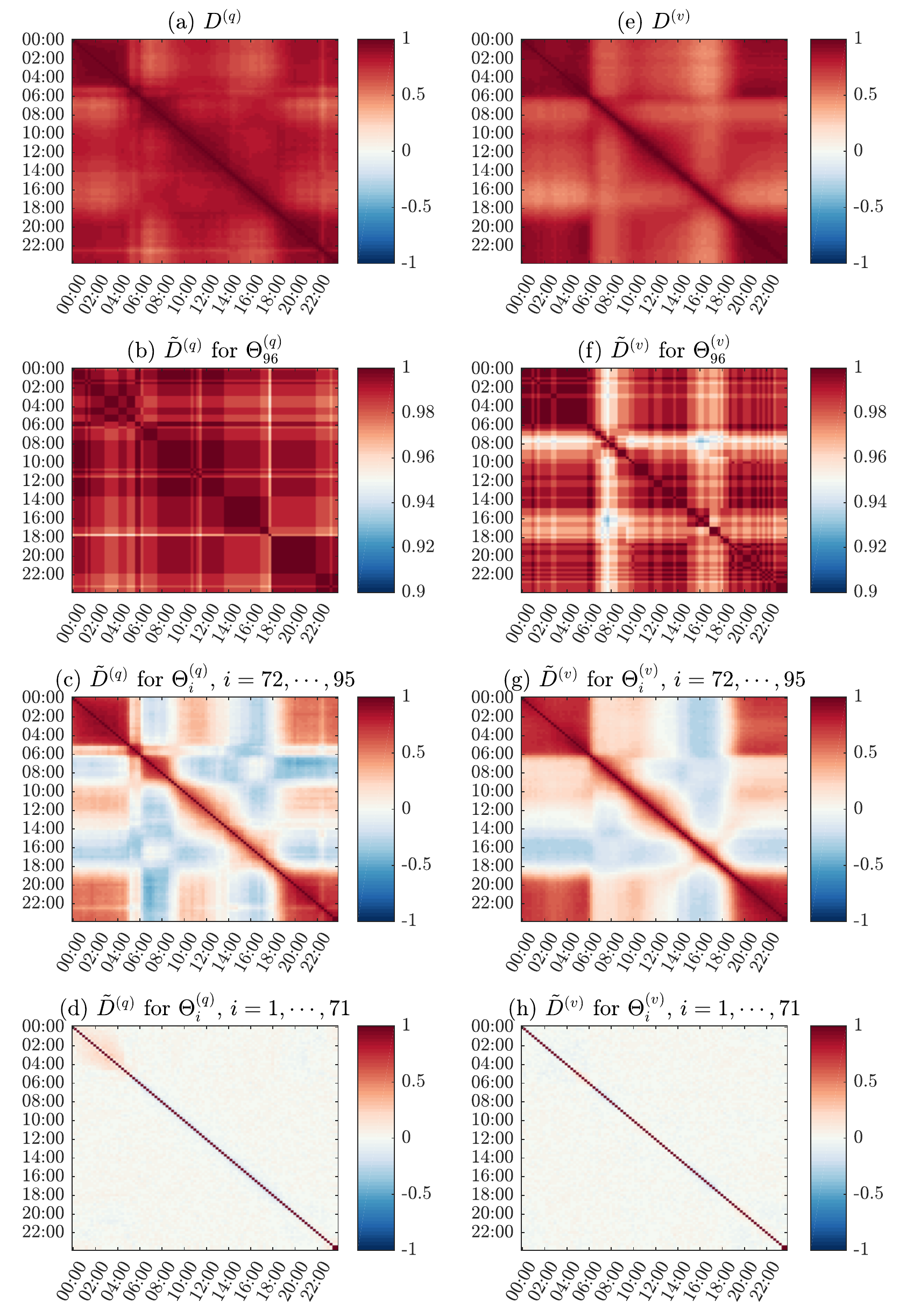}
\caption{The original correlation matrices (a) and (e) and the corresponding reduced-rank correlation matrices (b)--(d) and (f)--(h) for flows (a)--(d) and velocities (e)--(h), respectively. Each correlation matrix is first computed on daily basis and then averaged over a whole year. We use a different color scale for (b) and (f) to reveal their correlation structures.}
\label{fig3}
\end{center}
\vspace*{-0.5cm}
\end{figure}

Consider $K$ time series $G_k(t)$ of length $T$, where $t=1, 2, 3, \cdots , T$ corresponds to the first, second, third, ... , last 15 min-period from 00:00 to 23:59. These time series are the rows of the rectangular data matrix $G$ with dimensions $K\times T$ 
 \begin{equation} 
G=\left[\begin{array}{cccc}
G_{1}(1) & G_{1}(2) & \cdots & G_{1}(T) \\
G_{2}(1) & G_{2}(2) & \cdots & G_{2}(T) \\
 \vdots &  \vdots & \ddots & \vdots \\ 
G_{K}(1) &  G_{K}(2) & \cdots & G_{K}(T) \end{array}\right] \ .
\label{eq3.1.1}
\end{equation}
The columns of $G$ are the position series which capture the signals at different positions at the same time. Here, we are interested in these position series. We normalize each position series across each column to a zero mean and a unit standard deviation by
\begin{equation}
M_{k}(t)=\frac{G_{k}(t)-\langle G_{k}(t) \rangle_K}{\sqrt{\langle G_{k}(t)^2\rangle_K-\langle G_{k}(t)\rangle_K^2}} \ ,
\label{eq3.1.2}
\end{equation} 
resulting in a normalized $K\times T$ data matrix
\begin{equation} 
M=\left[\begin{array}{cccc}
M_{1}(1) & M_{1}(2) & \cdots & M_{1}(T) \\
M_{2}(1) & M_{2}(2) & \cdots & M_{2}(T) \\
 \vdots &  \vdots & \ddots & \vdots \\ 
M_{K}(1) &  M_{K}(2) & \cdots & M_{K}(T) \end{array}\right] \ .
\label{eq3.1.3}
\end{equation}
Here and hereafter, we indicate averages with respect to positions and times by $\langle \cdots \rangle_{K}$ and $\langle \cdots \rangle_{T}$, respectively. Our data consists of flows $q_k(t)$ and velocities $v_k(t)$. In the following, the superscript $(q)$ indicates a quantity of flows and the superscript $(v)$ indicates a quantity of velocities. A notation without any superscript indicates a general case. Because of the number of detectors across the road sections and the previously explained averaging over time intervals, we have $K=35$ as the number of road sections and $T=96$ as the time steps per day. In terms of the matrix $M$, the $T\times T$ correlation matrix of the position series, 
 \begin{equation}
D=\frac{1}{K}M^{\dag}M \ 
\label{eq3.1.4}
\end{equation}
reveals the correlations in time, i.e. the non-Markovian properties. We notice that $D$ does not have full rank, because $K=35<T=96$. The subscript $\dag$ indicates the transpose of a matrix. The structures of empirical correlation matrices $D$ for flows and velocities are masked by strongly positive correlations, as shown in figure~\ref{fig3} (a) and (e). To unravel the significant information contained in the correlation matrices, a spectrum decomposition is performed as follows.

\begin{figure}[tbp]
\begin{center}
\includegraphics[width=1\textwidth]{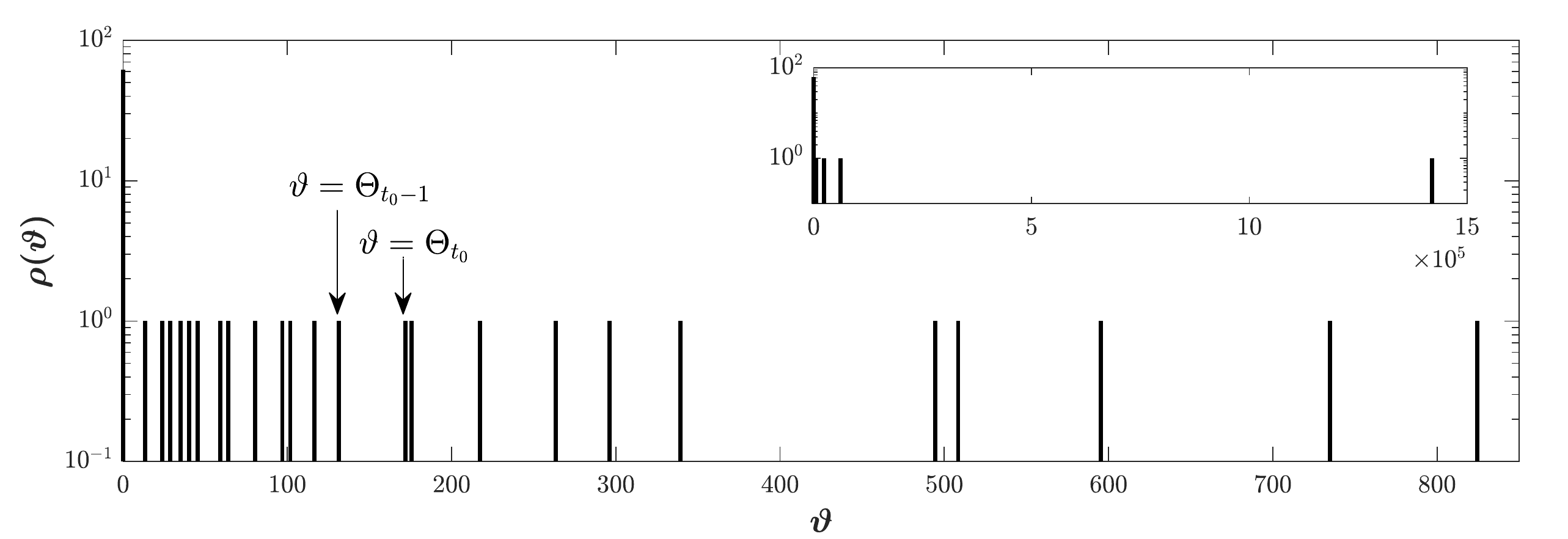}
\caption{The spectral density distribution $\rho(\vartheta)$ of eigenvalues $\vartheta$ from the original covariance matrix $\Sigma$ for the day of 1st January 2015. The horizontal axis is on a linear scale and the vertical axis is on a logarithmic scale. The inset plot displays the spectral density distribution of all eigenvalues.}
\label{fig2}
\end{center}
\vspace*{-0.5cm}
\end{figure}

By normalizing each position series of the data matrix $G$ across each column to the zero mean but not to the unit standard deviation, we obtain a new $K\times T$ data matrix $A$ with elements  
\begin{equation}
A_{k}(t)=G_{k}(t)-\langle G_{k}(t) \rangle_K \ ,
\label{eq3.1.5}
\end{equation}
which yields a $T\times T$ covariance matrix,
\begin{equation}
\Sigma=\frac{1}{K}A^{\dag}A \ .
\label{eq3.1.6}
\end{equation}
We apply a spectral decomposition to the covariance matrix,
\begin{equation}
\Sigma=V\Theta V^{\dag} \ ,
\label{eq3.1.7}
\end{equation}
where $\Theta$ is a diagonal matrix of eigenvalues in an ascending order
\begin{equation} 
\Theta=\mathrm{diag} (\Theta_1,\cdots,\Theta_T)
\label{eq3.1.8}
\end{equation}
and $V$ is a $T\times T$ orthogonal matrix whose columns are the corresponding eigenvectors $V(t)$ ($t=1, 2, \cdots, T$).
Thus, $\Theta_T$ is the largest eigenvalue and $\Theta_1$ the smallest eigenvalue. We write the covariance matrix $\Sigma$ as a sum of dyadic matrices
\begin{equation}
\Sigma=\sum_{t=1}^{T}\Theta_tV(t)V^{\dag}(t) \ ,
\label{eq3.1.9}
\end{equation}
which is the spectral expansion or decomposition. By considering only a part of this expansion, from $t=a$ to $t=b$, say,
\begin{equation}
\tilde{\Sigma}=\sum_{t=a}^{b}\Theta_tV(t)V^{\dag}(t) \ ,
\label{eq3.1.10}
\end{equation}
we can construct reduced-rank covariance matrices that reveal certain properties in a clearer fashion. In this way, we are able to extract three kinds of information, which are assumed to be the random noise with $a=1$ and $b=t_0-1$, the group motion with $a=t_0$ and $b=T-1$, and the collective motion of the whole system with $a=b=T$, similar to the information identified for financial markets~\cite{Plerou2002,Pharasi2019}. To give an example of how to choose $t_0$, we consider the spectral density
\begin{equation}
\rho(\vartheta)=\sum\limits_{t=1}^{T}\delta (\vartheta-\Theta_t) \ 
\end{equation}
of the eigenvalues of the covariance matrix $\Sigma$, displayed in figure~\ref{fig2}. With $T=96$ and $K=35$, it has $T-K+1=62$ zero eigenvalues and
$K-1=34$ non--zero ones. Even though this number is too small to justify a random matrix analysis, the Marcenko--Pastur eigenvalue density~\cite{Marchenko1967} can serve as a qualitative guideline. On the assumption of full randomness, the density has a bulk form where $\Theta_\mp=(\sqrt{T}\mp\sqrt{K})^2$ are the smallest and largest non--zero eigenvalues, which amount to $\Theta_-\approx 15$ and $\Theta_+\approx 247$. In the presence of correlations, this changes, and specific larger eigenvalues result, as seen in figure~\ref{fig2}. Nevertheless, there is an accumulation between $\Theta_-$ and $\Theta_+$ which is, due to the small number of eigenvalues, not yet developed into a bulk. Nevertheless, if we thus choose $t_0=72$ with a $\Theta_{t_0}$ slightly smaller than $\Theta_+$, we are sure to separate the spectral region revealing the correlation structure from the region governed by randomness.

The reduced-rank covariance matrix $\tilde{\Sigma}$ is well-defined, as will be explained in section~\ref{sec32}. Furthermore, we order the square roots of the diagonal elements in the reduced-rank covariance matrix $\tilde{\Sigma}$ in a diagonal matrix of standard deviations,
\begin{equation}
\tilde{\sigma}=\mathrm{diag} \left(\tilde{\sigma}_1,\cdots,\tilde{\sigma}_T \right)\ ,
\label{eq3.1.11}
\end{equation} 
and define the reduced-rank correlation matrix~\cite{Heckens2020}
\begin{equation}
\tilde{D}=\tilde{\sigma}^{-1} \tilde{\Sigma}  \tilde{\sigma}^{-1}\ .
\label{eq3.1.12}
\end{equation}
We always distinguish between the reduced-rank correlation matrix~\cite{Goldstein1997} and the original correlation matrix $D$. 

\begin{figure}[tb]
\begin{center}
\includegraphics[width=1\textwidth]{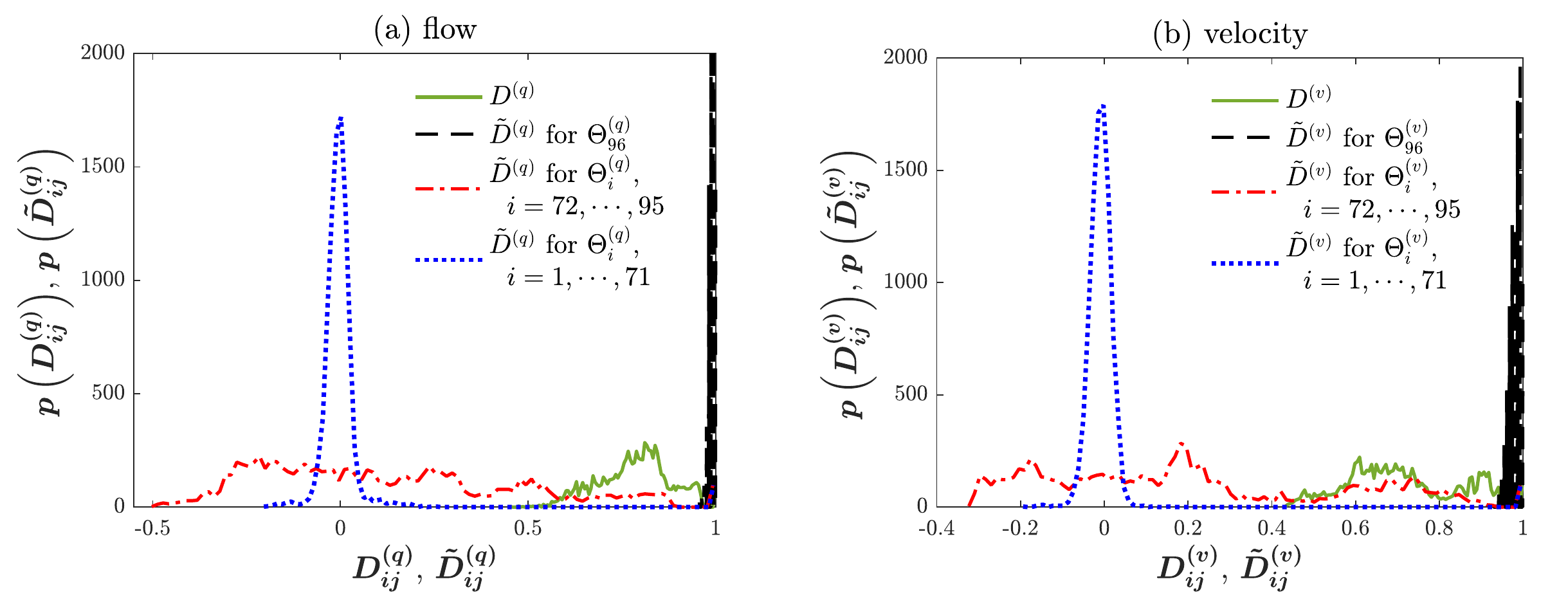}
\vspace*{0.1cm}
\caption{The probability density distributions of the elements $D_{ij}^{(x)}$ in the original correlation matrices and of the elements $\tilde{D}_{ij}^{(x)}$ in the reduced-rank correlation matrices, where $x=q$ for flows (a) and $x=v$ for velocities (b), respectively.}
\label{fig4}
\end{center}
\end{figure}

The three kinds of reduced-rank correlation matrices $\tilde{D}$ averaged over a whole year show remarkable differences in figure~\ref{fig3}. Only with the largest eigenvalues do the correlations in matrices $\tilde{D}$ approach one, implying that the collective motion of the whole system is almost completely correlated. Since strongly positive correlations are predominant,  the structural feature of $\tilde{D}$ in time is less distinct than the feature of $\tilde{D}$ with $\Theta_{72}$ to $\Theta_{95}$. The latter hints at information about group motion. If individual points in time are strongly correlated with the neighboring times, they form a group and exhibit a diagonal block feature in the matrix $\tilde{D}$. In all cases, the values in $\tilde{D}$ with $\Theta_{72}$ to $\Theta_{95}$ span a large range that contains both strongly positive and negative correlations, shown in figure~\ref{fig4}. Accordingly, the reduced-rank correlation matrix $\tilde{D}$ with group information displays rich structural features. In particular, diagonal blocks with strongly positive correlations in $\tilde{D}$ can be seen during the night, in the morning and afternoon rush hours, and also between the two rush hours. Although $\tilde{D}$ with the remaining small eigenvalues contains both positive and negative correlations as well, these correlations around zero are small except for the diagonal self-correlations. A lack of structural features in the correlation matrix $\tilde{D}$ with $\Theta_{1}$ to $\Theta_{71}$ makes the information very likely be noise. By comparison, $\tilde{D}$ with $\Theta_{72}$ to $\Theta_{95}$ is our best choice for identifying the states of the traffic system. Hereafter, the reduced-rank correlation matrix always indicates $\tilde{D}$ with $\Theta_{72}$ to $\Theta_{95}$.

\subsection{The role of the large eigenvalues}
\label{sec32}
A statistical analysis of 34 non-zero eigenvalues in the singular covariance matrix $\Sigma$ would not reliably uncover the role of the large eigenvalues. Instead, we return to the zero-mean data matrix $A$, which gives us a visualization of how the data behaves after a rank reduction. To this end, we carry out a singular value decomposition,
\begin{equation}
A=USV^{\dag} \ ,
\label{eq3.3.1}
\end{equation}
where $S$ is a $K\times T$ singular value matrix ($K<T$) with the elements $S_1,S_2,...,S_K$ in an ascending order,
\begin{equation} 
S=\left[\begin{array}{ccccccc}
S_1 	  & 0 		& \cdots & 0	&0 &\cdots &0	\\
0 	  & S_2 	& \cdots & 0	 &0 &\cdots &0	\\
\vdots & \vdots & \ddots & \vdots&\vdots & &\vdots	\\
0        &  0 	& \cdots & S_K	&0 &\cdots &0	\\
\end{array}\right] \ ,
\label{eq3.3.2}
\end{equation}
$U$ is a $K\times K$ orthogonal matrix with columns of the corresponding left eigenvectors $U(k)$ ($k=1, 2, \cdots, K$),  
and $V$ is a $T\times T$ orthogonal matrix with columns of the corresponding right eigenvectors $V(t)$ ($t=1, 2, \cdots, T$), which coincides with the eigenvector in section~\ref{sec31}. Replacing $A$ with equation~(\ref{eq3.3.1}), the original covariance matrix in equation~(\ref{eq3.1.6}) can be reformulated as
\begin{equation}
\Sigma=\frac{1}{K}VS^{\dag}SV^{\dag} \ .
\label{eq3.3.4}
\end{equation}
Since the covariance matrix $\Sigma$ is decomposed as according to equation~(\ref{eq3.1.7}), the relation between the singular value matrix $S$ and the eigenvalue matrix $\Theta$ reads
\begin{equation}
\Theta=\frac{1}{K}S^{\dag}S \ . 
\label{eq3.3.6}
\end{equation}
For each eigenvalue and the corresponding singular value, we therefore have
\begin{equation}
S_t=\sqrt{K\Theta_t} \ . 
\label{eq3.3.7}
\end{equation}
Similar to equation~(\ref{eq3.1.10}), the rank of the data matrix can be reduced by
\begin{equation}
\tilde{A}=\sum_{t=a}^{b}S_tU(t)V^{\dag}(t) \ .
\label{eq3.3.8}
\end{equation}
Using the orthogonal property of eigenvalues and equation~(\ref{eq3.3.7}), the reduced-rank covariance matrix $\tilde{\Sigma}$ can be  derived from the reduced-rank data matrix $\tilde {A}$ by
\begin{eqnarray}\nonumber
\tilde{\Sigma}&=&\sum_{t=a}^{b} \Theta_tV(t)V^{\dag}(t)  \\  \nonumber
&=& \frac{1}{K} \sum_{t=a}^{b} S_t^2V(t)U^{\dag}(t)U(t)V^{\dag}(t)  \\ \nonumber
&=& \frac{1}{K} \left(\sum_{t=a}^{b} S_tU(t)V^{\dag}(t)\right)^{\dag} \left(\sum_{t=a}^{b} S_tU(t)V^{\dag}(t)\right) \\  
&=& \frac{1}{K}\tilde{A}^{\dag}\tilde{A}\ .
\label{eq3.3.9}
\end{eqnarray}
Here, $\tilde{\Sigma}$ is a well-defined covariance matrix, since each position series of $\tilde{A}$ is normalized to the zero mean. To demonstrate this, we introduce a $K$ dimensional unit vector $e=(1,\cdots, 1)$ and a $p$ dimensional zero vector $\emptyset_p=(0,\cdots, 0)$ so that 
\begin{equation}
A^{\dag}e=\emptyset_T \qquad \mathrm{and} \qquad  AA^{\dag}e=\emptyset_K \ .
\label{eq3.3.10}
\end{equation} 
From the singular value decomposition~(\ref{eq3.3.1}), we find
\begin{equation}
A^{\dag}=\sum_{t=1}^{T}S_tV(t)U^{\dag}(t) \ 
\label{eq3.3.11}
\end{equation}
for the transpose. Since $K<T$, $A^{\dag}$ has $K-1$ non-zero singular values and one zero singular value. Equation~(\ref{eq3.3.11}) yields
 \begin{equation}
U^{\dag}(t)e=0  \ 
\label{eq3.3.12}
\end{equation} 
for all $t$ except one for which we have $S_t=0$. When the range of $t$ between $1$ and $T$ in equation~(\ref{eq3.3.11}) is narrowed to the range between $a$ and $b$ ($1\leq a\leq b\leq T$), the rank of $A^{\dag}$ is reduced to the one of $\tilde{A}^{\dag}$. Taking the above discussion into account, we can obtain
\begin{equation}
\tilde{A}^{\dag}e=\emptyset_T  \ ,
\label{eq3.3.13}
\end{equation}
which means each position series of $\tilde{A}$ is normalized to the zero mean. 

As a result, the role of the eigenvalues in the original covariance matrix can be traced back to the singular values in the zero-mean data matrix $A$. In other words, the contribution of the singular values from the zero-mean data matrix $A$ gives rise to the contribution of the corresponding eigenvalues in the covariance matrix $\Sigma$.

\begin{figure}[htbp]
\begin{center}
\includegraphics[width=1\textwidth]{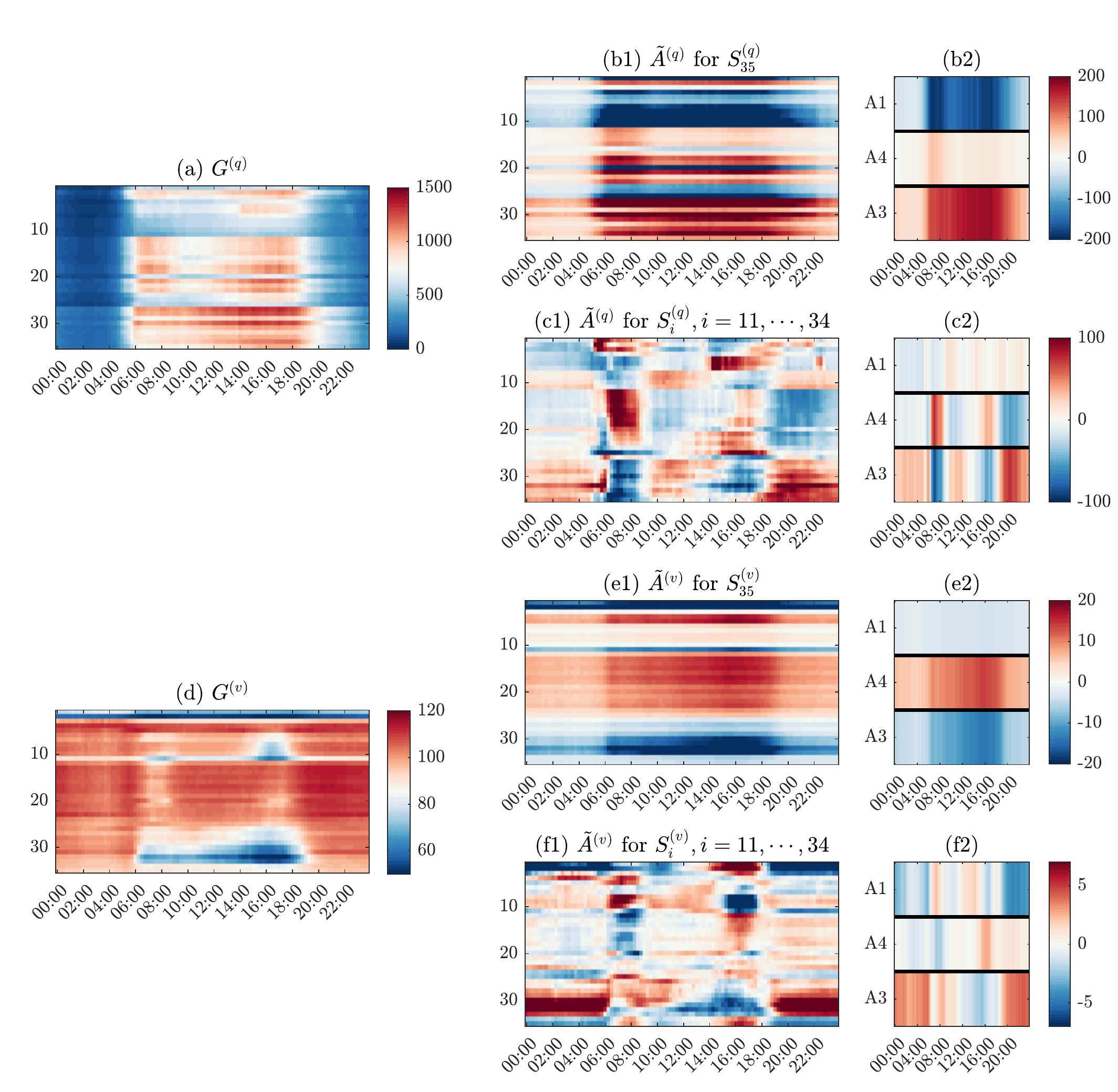}
\vspace*{-0.5cm}
\caption{ A comparison among $35\times 96$ original data matrices $G^{(x)}$ (a) and (d), reduced-rank data matrices $\tilde{A}^{(x)}$ with $S_{35}^{(x)}$ (b1) and (e1), and reduced-rank data matrices $\tilde{A}^{(x)}$ with $S_{i}^{(x)}, i=11,\cdots, 34$ (c1) and (f1), where $x=q$ for flows (a)--(c) and $x=v$ for velocities (d)--(f). The labels on the vertical axes indicate the sections and those on the horizontal axes indicate time. For each time step, averaging $\tilde{A}^{(x)}$ with $S_{35}^{(x)}$ in (b1) and (e1), respectively, across the sections in each motorway, i.e., the Cologne orbital motorways A1 (sections 1–11), A4 (sections 12–25) and A3 (sections 26–35), results in (b2) and (e2). For each time step, averaging $\tilde{A}^{(x)}$ with $S_{i}^{(x)}, i=11,\cdots, 34$ in (c1) and (f1), respectively, across the sections in each motorway results in (c2) and (f2). The motorways A1, A4 and A3 are separated by black lines.}
\label{fig5}
\end{center}
\end{figure}

Therefore, to figure out the role of the largest eigenvalue, we can resort to the reduced-rank data matrices $\tilde{A}$ with the largest singular value $S_{35}$, shown in figure~\ref{fig5}. Compared to the original data matrices $G$, the matrices $\tilde{A}$ with $S_{35}$ show a feature which is that sections depend on motorways more than on time. Put differently, the sections present distinct behaviors on motorways but collective behaviors in time. The specific feature in time dominates in the matrices $\tilde{A}$ with other large singular values $S_{i}, i=11,\cdots, 34$ in figure~\ref{fig5}. In this case, the typical features of morning and afternoon rush hours appear. Consecutive times with similar features can be grouped together, leading to a group behavior in time. Accordingly, the largest eigenvalue reveals the collective information in time, while other large eigenvalues reveal the group information in time.

In figure~\ref{fig5} (d), we notice a very low speed for section 2 all the time. Such a section, however, does not influence the reduced-rank correlation matrices, as we remove the largest eigenvalue from the original correlation matrices. The largest eigenvalue, as aforementioned, describes the collective information in time, for instance, for section 2, a speed limit is relevant which is active on the bridge over the Rhine in Leverkusen~\cite{bridge}, which lies directly behind section 2. Obvious evidence for this can be found in figure~\ref{fig5} (f1), where the collective information of section 2 is wiped out and the group information depending on time is evident especially during the daytime. If any section-dependent information remains after removing the largest eigenvalue, the average in the temporal correlation function~(\ref{eq3.1.4}) or in the temporal covariance function~(\ref{eq3.1.6}) is performed over different sections so as to dilute the influence of a specific section. This is the advantage of using the temporal correlation instead of the spatial correlation.

\section{Quasi-stationary states for traffic systems}
\label{sec4}

Clustering is a widely used technique in statistical data analysis for grouping a set of objects into different clusters, so that the objects in the same cluster are more similar to each other than to those in other clusters. To describe the motivation for our approach, we provide some technical comments on commonly used clustering methods in section~\ref{sec41}. Considering all reduced-rank correlation matrices in 2015 as a set of our objects, we group them in different clusters with a centroid-based clustering method~\cite{Sun2014}, i.e., $k$-means clustering~\cite{Lloyd1982}, in section~\ref{sec42}. We refer to each cluster as a quasi-stationary state and then characterize the five quasi-stationary states identified by clustering in section~\ref{sec43}.

\subsection{Introduction to clustering methods}
\label{sec41}

Despite the existence of more than 100 clustering methods, most of which have limited applications, there are four commonly used clustering types~\cite{Kameshwaran2014,Xu2015}: hierarchical clustering~\cite{Johnson1967,Murtagh1983}, centroid-based clustering~\cite{Sun2014}, distribution-based clustering~\cite{Xu1998} and density-based clustering~\cite{Kriegel2011}. Hierarchical clustering is further categorized into agglomerative (bottom-up) clustering and divisive (top-down) clustering. Agglomerative clustering means that single clusters recursively merge into the most similar clusters, and divisive clustering indicates that a single cluster containing all the data points recursively splits into the most appropriate clusters. Hierarchical clustering does not require {\it a priori} information about the number of clusters, but conversely it is difficult to control and correctly identify the number of clusters, especially for non-hierarchical data. The complexity of this algorithm, as well as its low efficiency when used with iterative computation also reduces its popularity to some extent. Centroid-based clustering, e.g., $k$-means clustering~\cite{Lloyd1982},is a method often used to classify objects into their nearest centroids, where the number of centroids $k$ is preset and the initial $k$ centroids are selected randomly. Despite the disadvantage that the number of clusters is required, this type of clustering algorithm has relatively low complexity and high computing efficiency. Distribution-based clustering, e.g. Gaussian mixture model clustering~\cite{Rasmussen2000}, assumes that data follow certain distributions. Thus, data with the same distribution belong to the same cluster. This gives the probability that a data point belongs to a distribution or a cluster, but it may cluster incorrectly if it suffers from overfitting due to many parameters or if the type of distribution of the given data is unknown. Density-based clustering, e.g. density-based spatial clustering of applications with noise~\cite{Ester1996}, ordering points (to identify the clustering structure)~\cite{Ankerst1999} and mean shift~\cite{Comaniciu2002}, identifies the data points in the region with high point densities as the same cluster~\cite{Kriegel2011}. In contrast, the data points in the low-density regions (as distinct from the high-density region) are considered to be noise and outliers. This type of clustering can be arbitrarily shaped in the data space and is highly efficient at clustering. However, it has difficulties in clustering data of varying density and high dimensions and fails to assign the outliers to clusters.

\subsection{Clustering with $k$-means algorithm}
\label{sec42}

Despite the pros and cons of the many clustering methods introduced in section~\ref{sec41}, almost all of the methods require some parameters to be preset. For instance, a distance threshold for the hierarchical clustering~\cite{Johnson1967,Murtagh1983}, the distribution parameters for the distribution-based clustering~\cite{Xu1998}, a density criterion for the density-based clustering~\cite{Kriegel2011}, etc. are required to be preset to obtain an ideal number of clusters that should be neither too many nor too few. These parameters, however, are more difficult to determine than an integer for the count of clusters. Here, considering the low complexity and high computing efficiency in contrast to the other clustering methods introduced in section~\ref{sec41}, we use the $k$-means clustering, which can be simply described as follows: 
\begin{itemize}[leftmargin=0.6cm]
\setlength\itemsep{-4pt}
\item[(a)] Define a metric and a distance between observations.
\item[(b)] Choose $k$ initial centroids at random for all observations. 
\item[(c)] Compute the distances between each observation and each centroid. 
\item[(d)] Assign each observation to the cluster of its closest centroid and re-label all observations with the indices of clusters. 
\item[(e)] Compute the average of all observations in each cluster to find $k$ new cluster centroids. 
\item[(f)] Repeat steps (c)--(e) until the labels of all observations do not change or iterations reach the preset maximal number. 
\end{itemize}

To find quasi-stationary states with correlation matrices of position series in traffic systems, as opposed to the states with correlation matrices of time series that are required for financial markets~\cite{Munnix2012,Guhr2015,Chetalova2015a,Chetalova2015b,Rinn2015,Stepanov2015}, we need to cluster the reduced-rank correlation matrices for a whole year. The reduced-rank correlation matrix for day $i$ is $\tilde{D}(i)$ with elements $\tilde{D}_{t_kt_l}(i)$ for two times $t_k$ and $t_l$. Furthermore, we define a similarity measure to quantify the difference between the correlation structures of two days,
\begin{equation}
\eta_{ij}=\big\langle | \tilde{D}_{t_kt_l}(i)-\tilde{D}_{t_kt_l}(j) |\big\rangle_{t_kt_l} \ ,
\label{eq3.4.1}
\end{equation}
where $|\cdots|$ is the absolute value and $\langle \cdots \rangle_{t_kt_l}$ stands for the average over all matrix elements. All similarity measures are entered into an $N\times N$ similarity matrix $\eta$, where $N$ is the number of reduced-rank correlation matrices. To proceed with the clustering, we may interpret the rows of $\eta$ as observations, i.e. the reduced-rank correlation matrices, and the columns as variables, i.e. the measured similarity between two correlation structures. The distance between two observations is defined as a squared Euclidean distance
\begin{equation}
d_{ij}=\sum\limits_{n=1}^{N}(\eta_{in}-\eta_{jn})^2 \ ,
\label{eq3.4.2}
\end{equation}
where $N=362$. If two matrices have common similarities with other matrices, the two matrices very probably have similar structures as well as a short distance between them. Given the squared Euclidean distance matrix, the $k$-means clustering can be carried out. 

\begin{figure}[t]
\begin{center}
\includegraphics[width=1\textwidth]{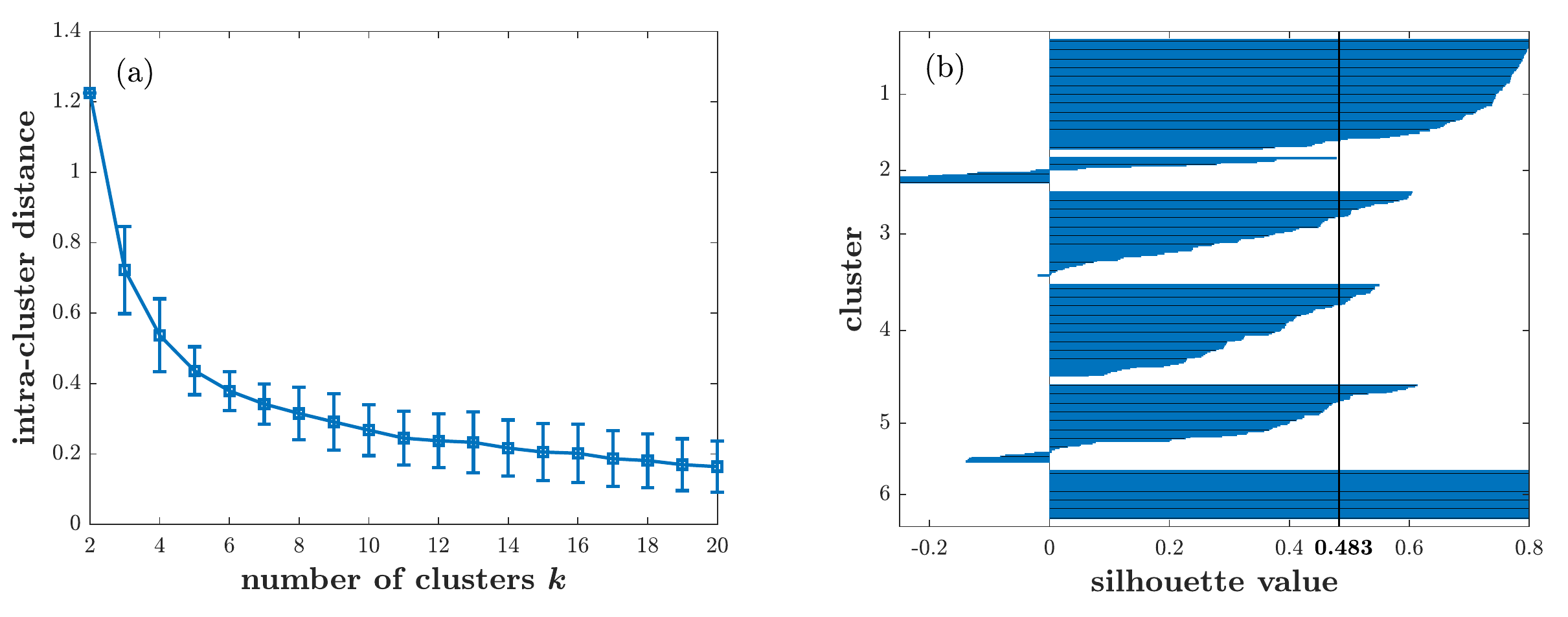}
\vspace*{-0.5cm}
\caption{(a) The intra-cluster distance, i.e. averaged squared Euclidean distance of all the points within clusters, versus the number of clusters $k$, where the error bar represents the standard deviation of the intra-cluster distances resulting from 500 runs of clustering with 500 randomly assigned initial centroids. For $k\geq 3$ the minimum error bar appears at $k=6$. (b) Silhouette values for five clusters, where the averaged silhouette value is 0.483.}
\label{fig6}
\end{center}
\end{figure}

To determine the optimal number of clusters $k$, we employ the following method proposed by Ref.~\cite{Pharasi2018}. For each given $k=1,\cdots, 20$, the $k$-mean clustering is run 500 separate times with 500 initial assignments of centroids at random. Each initial assignment of centroids is a $k\times N$ matrix of centroid starting locations. Each row of the matrix corresponds to one of $k$ initial cluster centroid positions. In each run, the squared Euclidean distance between points and the centroid of the cluster that the points belong to can be calculated. The squared Euclidean distances for all clusters are averaged, resulting in an intra-cluster distance. An average of intra-cluster distances for 500 runs and their standard deviation can be worked out. For distinct clusters, the different initial assignments of centroids will yield similar results, which lead to a small standard deviation of intra-cluster distances. On the contrary, for close clusters or clusters with overlapping positions, different initial assignments of centroids greatly affect the final results, giving rise to a large standard deviation. Therefore, a small standard deviation suggests distinct clusters with the optimal number of clusters. We restrict the number of clusters to $3\leq k\leq 20$. In this range, a small standard deviation can be found at $k=6$ from the error bars around the average intra-cluster distances in figure~\ref{fig6} (a). 

For 2015, there are 362 available reduced-rank correlation matrices, i.e. 362 observations, which are classified into six clusters using the $k$-means algorithm, where the sixth cluster contains the observations with missing data for the corresponding day. The missing data may be caused by defective detectors, construction sites, road closures, problems with the service providing the data, or for other reasons. To evaluate the performance of the clustering, we test the results with silhouette values, which measure how close an observation is to the observations in its own cluster and how far it is from the observations in other clusters~\cite{Kaufman2009}. The resulting value ranges from -1 to +1, where a high positive value indicates that an observation is appropriately classified in its own cluster rather than in the neighboring clusters. From figure~\ref{fig6} (b), we find that most of the observations are classified appropriately into their own clusters, especially for the observations in the first, third, fourth and sixth clusters. The average of the silhouette values reaches 0.483, implying a good clustering performance on the whole.

\begin{figure}[h!]
\begin{center}
\includegraphics[width=0.98\textwidth]{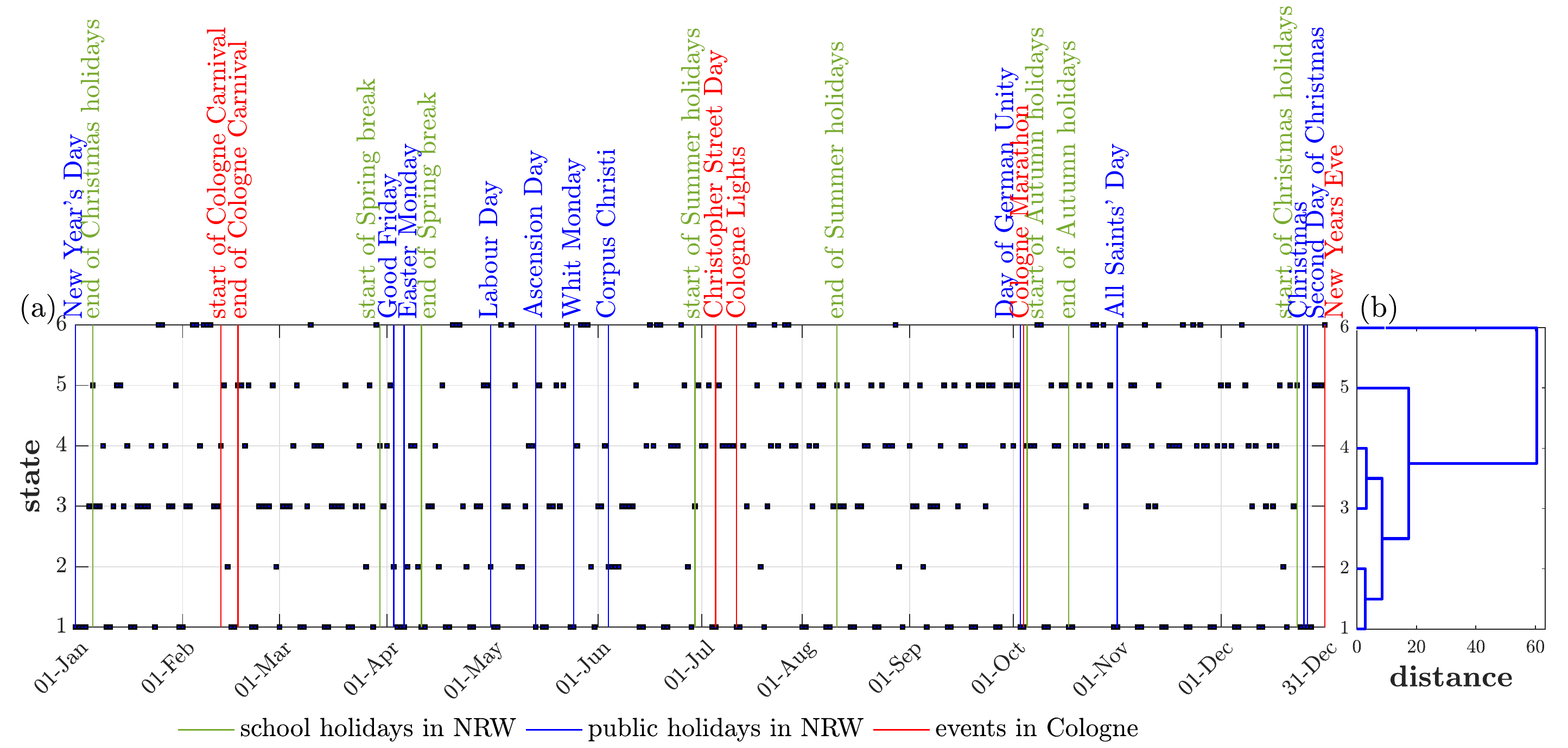}\vspace{0.3cm}
\includegraphics[width=1\textwidth]{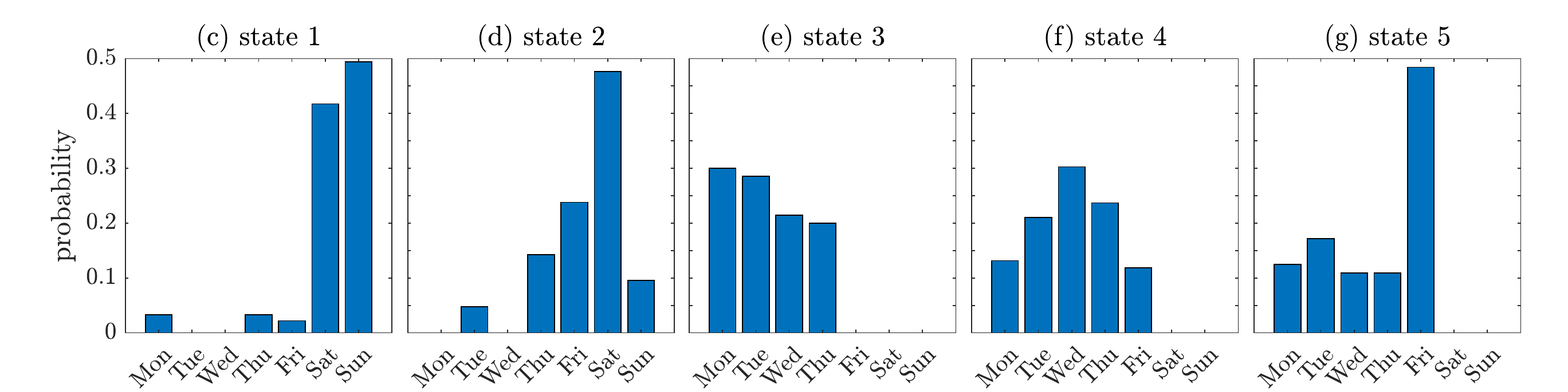}
\caption{(a) Time evolution of the traffic quasi-stationary states shown in black dots, where the days with missing data are classified as state 6. The colored lines indicate the states emerging on holidays or days with major events. (b) The corresponding hierarchical tree of six states resulting from the agglomerative hierarchical clustering using the metric of squared Euclidean distance, where the missing values of state 6 are set to zero. (c)--(g) The probability distribution for every weekday on which a given state occurs.}
\label{fig7}
\end{center}
\end{figure}

\subsection{Features of states identified by clustering}
\label{sec43}

Since each cluster composed of many similar reduced-rank correlation matrices hints at a universal traffic condition, we refer to each cluster as a quasi-stationary state for the traffic system under investigation. The time evolution of the states is shown in figure~\ref{fig7} (a), which is compared with regional holidays and events. Hence we get a rough outline that (1) the majority of public holidays in North Rhine-Westphalia (NRW) (the state that Cologne belongs to) and the majority of events in Cologne occur either in state 1 or state 2; (2) during school holidays, state 3 emerges infrequently; (3) state 1 is constant across the whole year while states 3 and 4 are frequently present during the first and the second half years, respectively. The transition from state 3 as the frequently occurring state during the first half of the year to state 4 as the frequently occurring state during the second half of the year is very probably due to construction that added lanes to the A3 motorway between Cologne-M\"ulheim and Leverkusen~\cite{news}. The roadwork started in June 2015 and had an impact on the traffic. Points (1) and (2) above lead us to take a closer look at how the states are related to the type of day. Figures~\ref{fig7} (c)--(g) display the probability distribution for every weekday on which a given state occurs. States 1 and 2 have a high probability of appearing on a Saturday or a Sunday, while states 3, 4 and 5 only emerge on workdays, and state 5 is more likely to show up on a Friday. These probability distributions provide a good explanation of why states 1 and 2 are close and states 3 and 4 are close in the hierarchical tree in figure~\ref{fig7} (b). Since the sixth cluster contains the observations with missing data across up to a half or a whole day, we focus our attention on the first five states in the following.

\begin{figure}[tbp]
\begin{center}
\raggedleft
\includegraphics[width=1\textwidth]{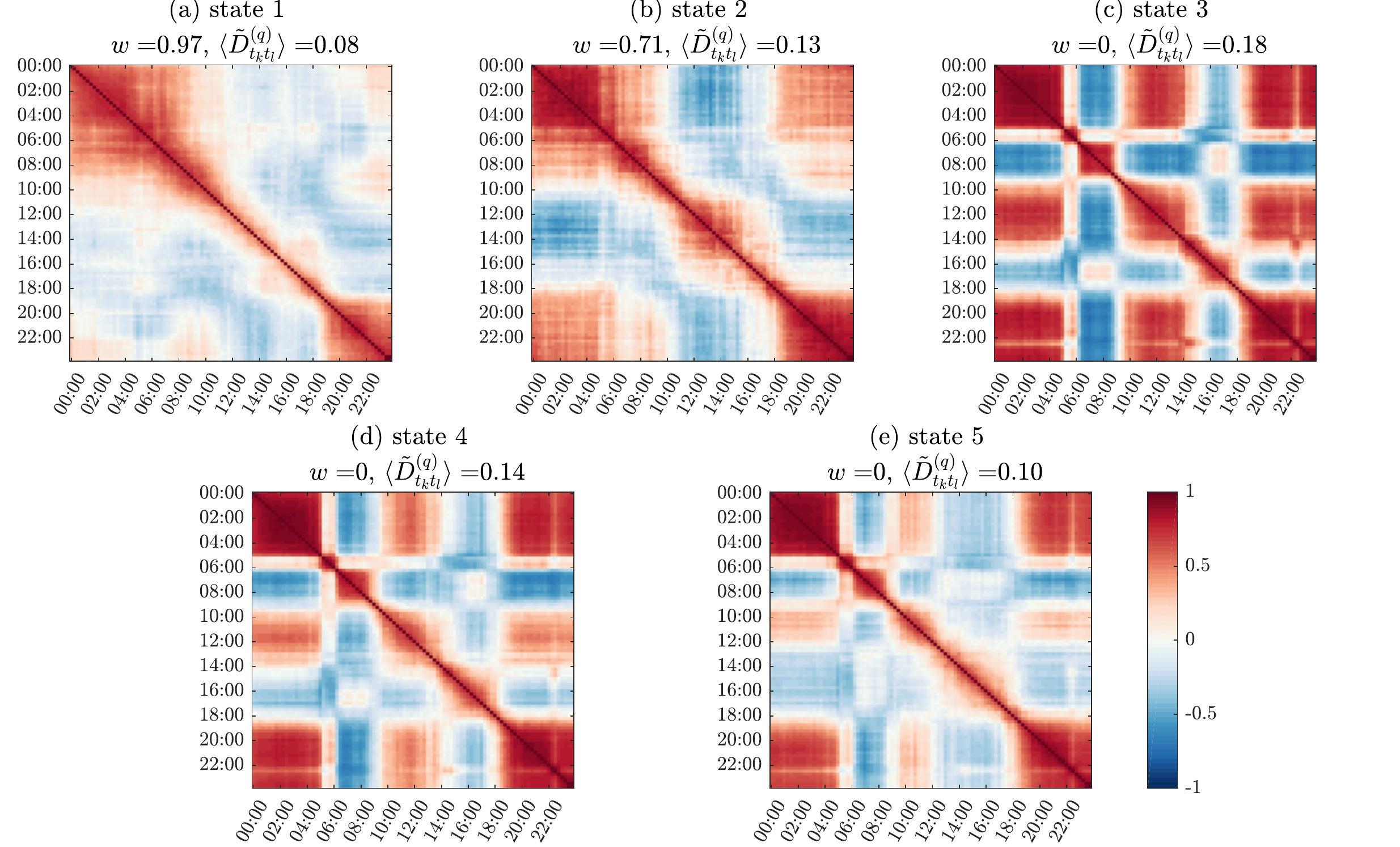}
\caption{The averaged reduced-rank correlation matrices of flows for five states, where $w$ is the proportion of holidays (including weekends and public holidays in NRW) to total days in each cluster, and $\langle \tilde{D}_{t_kt_l}^{(q)}\rangle$ is the average of all elements in each matrix.}
\label{fig8}
\end{center}
\end{figure}

To obtain an averaged reduced-rank correlation matrix for a certain state, we average all reduced-rank correlation matrices belonging to this state. The average extracts the essential and universal features of the correlation structures while ruling out the uncommon features caused by occasional events on each day. The averaged reduced-rank correlation matrices for five states are shown in figure~\ref{fig8}, where $w$ is used to quantify the proportion of holidays (including weekends and public holidays in NRW) to total days in each state. In each correlation structure, we mainly focus on the diagonal blocks with strongly positive correlations during daytime between 6:00 and 19:00, as these diagonal blocks reveal strongly correlated groups of time.

State 1 composed of $97\%$ holidays reflects a typical correlation pattern for holidays, and is called the holiday state, where diagonal blocks with sharp boundaries during daytime are absent in its correlation structure. In contrast, the diagonal blocks are distinct in the correlation structures of states 3, 4 and 5 that only contain workdays, known as workday states. The diagonal blocks with strongly positive correlations across sequential points in time imply that a free (or congested) flow is more likely to be followed by another free (or congested) flow. During rush hours, there is a higher probability of finding a congested flow instead of a free flow. Therefore, the diagonal block around rush hours hints at a persistent congested flow. In contrast, outside rush hours, the congestion case is relieved and a free flow is very likely to be found, so that the diagonal block around non-rush hours implies a persistent free flow. For both cases, the larger the block, the longer the congested or free flow persists. For state 3, the diagonal blocks in the correlation structures are remarkable both during rush hours and non-rush hours. From state 4 to state 5, the blocks during afternoon rush hours and during non-rush hours between 10:00 and 15:00 become smaller and smaller. Meanwhile the correlations become weaker and weaker. The difference between states 3 and 4 versus state 5 is due to Fridays, which dominate in state 5, as shown in figures~\ref{fig7} (e)--(g). As Friday bridges the other four workdays and the weekend, people rush to work in the morning but may finish work at a different time from the finishing time of the first four days, due to the coming weekend. This may explain the feature of the strongly positive correlations during morning rush hours but comparatively weak correlations in the following intraday time. State 2, comprised of $71\%$ holidays and $29\%$ workdays, shows a mixed feature that combines the correlation structures of holiday states with the diagonal blocks found in the correlation structures of the workday states. Thus, it is called the mixed state.

\section{Projections of quasi-stationary states}
\label{sec5}

To understand the five quasi-stationary states in a deeper fashion, we map them onto other related matrices in the following way. For each day, we work out a reduced-rank correlation matrix of flows, a reduced-rank correlation matrix of velocities and a matrix of traffic states introduced in section~\ref{sec52}. Based on the clustering described in section~\ref{sec4}, we assign a state to each reduced-rank correlation matrix of flows. That means we assign each day a state. The reduced-rank correlation matrix of velocities and the matrix of traffic states share the same state with the corresponding day. We then average over all reduced-rank correlation matrices of velocities (all matrices of traffic states) with the same state to obtain an averaged reduced-rank correlation matrix (matrix of traffic states) for a given state. In this way, we map the five quasi-stationary states onto velocity correlation matrices in section~\ref{sec51} and onto traffic states both in time and space in section~\ref{sec52}.

\subsection{Mapping five states onto velocity correlation matrices}
\label{sec51}

\begin{figure}[tbp]
\begin{center}
\raggedleft
\includegraphics[width=1\textwidth]{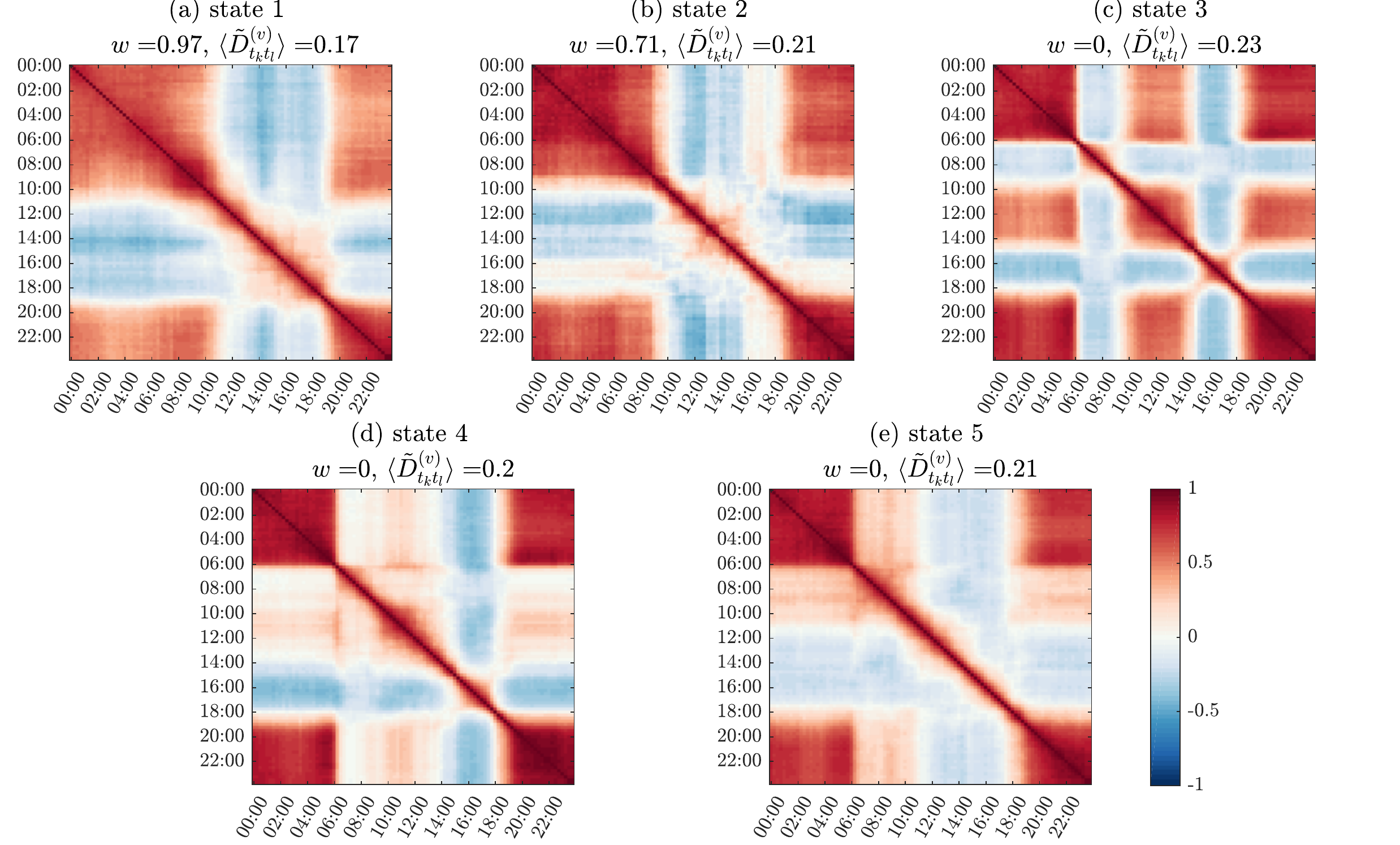}
\caption{The reduced-rank correlation matrices of velocities for five states that are identified with the reduced-rank correlation matrices of flows, where $w$ is the proportion of holidays (including weekends and public holidays in NRW) to total days in each cluster, and $\langle \tilde{D}_{t_kt_l}^{(v)}\rangle$ is the average of all elements in each matrix.}
\label{fig9}
\end{center}
\end{figure}

To dissect the five states further, we map them onto the averaged reduced-rank correlation matrices for velocities, displayed in figure~\ref{fig9}. The diagonal block from 19:00 to 10:00 (state 1) or 9:00 (state 2) of the next day reveals strongly positive correlations between high velocities, as the flow is low during these holiday periods. In contrast, during holiday lunch times, with an increase in flows, the velocities decrease. The diagonal block shown at this time period very probably implies strongly positive correlations between low velocities. Since state 2 contains a small number of workdays, small diagonal blocks with a width of 2 hours can be found as well. For workday states, the diagonal block during rush hours indicates strongly positive correlations of low velocities while during non-rush hours, it indicates strongly positive correlations of high velocities. The larger the block, the longer the high (or low) velocity remains and the free (or congested) case persists. During daytime, the diagonal blocks are significant in state 3 but become more and more blurred and even disappear from state 4 to state 5.

\subsection{Mapping five states onto traffic states}
\label{sec52}

\begin{figure}[tb]
\begin{center}
\includegraphics[width=1\textwidth]{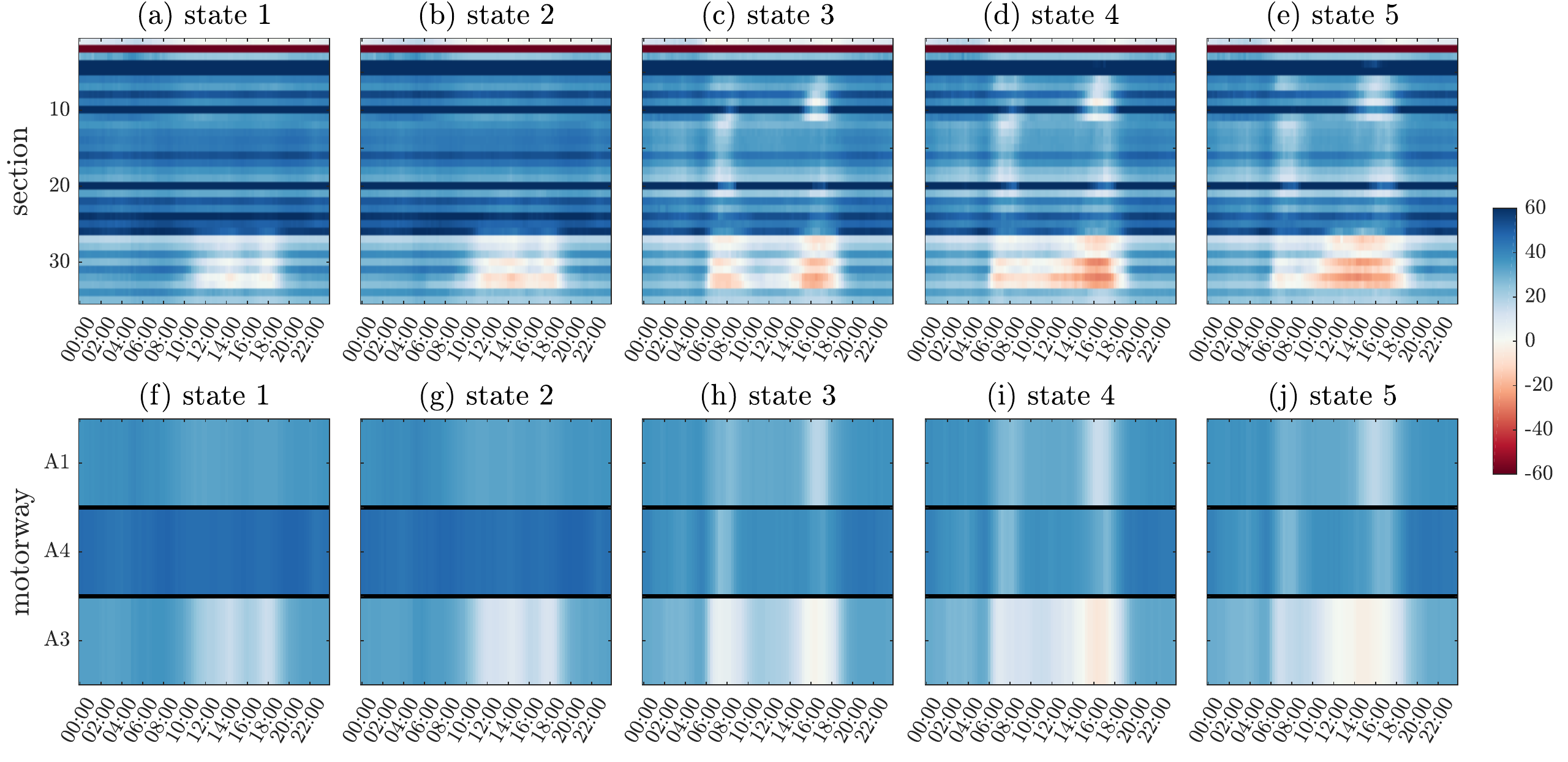}
\vspace*{-0.5cm}
\caption{(a)--(e) The traffic states shown in $35\times 96$ data matrices of $\Delta v_i(t)$ for five states that are identified with the reduced-rank correlation matrices of flows. (f)--(j) At each time step of each state, averaging $\Delta v_i(t)$ over all the sections in each motorway (A1, A3 or A4) results in an average traffic state of the corresponding motorway, where A1, A4 and A3 are separated by black lines. Blue indicates a free traffic state while red indicates a congested traffic state. White means that the vehicle speed is close to the critical velocity of each section that distinguishes free and congested flows.}
\label{fig10}
\end{center}
\end{figure}

To understand the reason why the five states, in particular the three workday states, behave differently, we map the five quasi-stationary states onto time-dependent traffic states~\cite{Seo2017}. As it is impossible for different sections to have exactly the same traffic conditions or environmental factors, for instance, speed limits, we use the critical velocity of each section to distinguish congested flows from free flows. Here, the critical velocity is defined as the ratio of the maximal measured flow within one year to the corresponding density~\cite{Kerner2012}
\begin{equation}
v_i^{(0)}(t)=\frac{q_{i,\mathrm{max}}^{(\mathrm{free})}(t) }{\rho_{i,\mathrm{max}}^{(\mathrm{free})}(t) } \ ,
\label{eq5.2.1}
\end{equation}
where $t$ is a time step with an interval of 15 min. A traffic state with a velocity above (below) the corresponding critical velocity is defined as a free (congested) state. Thus, for a free (congested) state, the difference between a velocity and its corresponding critical velocity, i.e., $\Delta v_i(t)=v_i(t)-v_i^{(0)}(t)$, is a positive (negative) value. The magnitude of $|\Delta v_i(t)|$ indicates the degree of the free or congested state. For each day, we have a matrix of $\Delta v_i(t)$ where the rows and columns indicate 35 sections and 96 time steps, respectively. Averaging over all these matrices for each quasi-stationary state eliminates the disturbance caused by occasional events and leads to the traffic states shown in figure~\ref{fig10}, which is analyzed as follows.

Due to a damage-induced speed limit on the Rhine bridge in Leverkusen~\cite{bridge}, section 2 (leading to the bridge) is heavily congested for all states. Except for this section, we barely find any congestion for sections on the A1 and A4 motorways. The velocities for the two traffic states decrease on the A3 motorway and fall to around the critical velocities between 10:00 and 19:00. During this time period, there is basically little congestion for state 1 but it can be seen in several sections of the A3 within state 2. In contrast to the sections of the A1 and A4 motorways, some sections of the A3 are less capable of resolving congestion because of the typical traffic behaviors on workdays included in state 2. For three workday states, a decrease in velocities is visible during the morning and afternoon rush hours. This decrease does not lead to heavy congestion for sections of the A1 and A4, but it affects the sections on the A3 strongly. On the A3 motorway, the congestion in the morning and afternoon rush hours is separated by free flows between these times in state 3. However, the congestion from the morning rush hour is not resolved until the end of the afternoon rush hour for several sections in states 4 and 5. Usually, between the morning and afternoon rush hours, the free flows produce strongly positive correlations in time, resulting in the diagonal blocks shown in figures~\ref{fig8} (c) and \ref{fig9} (c). When congestion occurs during this time period, the correlations of the free flows are diminished. That is the reason why the diagonal blocks in the correlation structures of flows or velocities gradually become blurred in figures~\ref{fig8} (d) and (e) and \ref{fig9} (d) and (e). Furthermore, from states 3 to state 5, the time period for which the afternoon rush hour is in progress gradually increases in width. This means that the vehicle flows that are concentrated in a small time period with a strong speed dependence on neighboring vehicles are spread over a larger time period with a comparatively weak speed dependence between vehicles. As a result, diagonal blocks with strongly positive correlations are weakened when the time period of the afternoon rush hour is extended.

\section{Conclusions}
\label{sec6}

Identifying states with correlation structures for a traffic system is important to understand traffic features both in space and time. To this end, we worked out the correlation matrix of traffic flows. This was done by studying covariance and correlation matrices of the position series in contrast to the usual approach involving the time series. Hence, the covariance and correlation matrices worked out here give information on the various non-Markovian features of traffic. To figure out the essential information hidden in the correlation matrix, spectrum filtering was carried out, resulting in a reduced-rank correlation matrix with specific eigenvalues, where the largest eigenvalue revealed collective time information, the other large eigenvalues revealed group time information, shown as diagonal blocks in a correlation matrix, and the small eigenvalues contained noise information.

Since the reduced-rank correlation matrix with the large eigenvalues subsequent to the largest one exhibits a distinct structural feature depending on time, we considered it as a candidate for clustering with the $k$-means method. We classified the 362 available reduced-rank correlation matrices for 2015 into six clusters, where the 6th cluster was for the matrices with a lot of missing data. The other five clusters, called quasi-stationary states, present three types, the holiday state (state 1), the workday states (states 3--5) and a mixed state of holidays and workdays (state 2). The workday states exhibit diagonal blocks in their correlation structures that indicate strongly correlated groups of time, e.g. the correlations of congested flows during rush hours and the correlations of free flows during non-rush hours. From state 3 to state 5, the diagonal blocks become more and more blurred after the morning rush hour. Due to the inclusion of workdays, state 2 also displays strongly correlated groups between the two rush hours. For state 1, the strongly correlated group is not so obvious during the daytime.

We projected the five states onto reduced-rank correlation matrices for velocities. Regardless of the small difference, the correlation structures for velocities exhibit similar features to those for flows in general. We further reflected the five states onto traffic states, whereby a free or congested state was revealed in both space and time. The change of congested states after the morning rush hour on the A3 motorway explains the gradually blurred diagonal blocks from state 3 to state 5. The mapping from quasi-stationary states onto free or congested traffic states helps to identify traffic jams and environmental factors, e.g. construction sites. With a series of historical quasi-stationary states, a predication for future states, either future quasi-stationary states or future traffic states, is possible using methods such as machine learning. In this sense, some management measures based on a precursor in traffic states could be made in advance to improve traffic efficiency.

\section*{Acknowledgements}

We gratefully acknowledge funding via the grant ``Korrelationen und deren Dynamik in Autobahnnetzen'', Deutsche Forschungsgemeinschaft (DFG, 418382724). We thank Strassen.NRW for providing the empirical traffic data. We also thank Sebastian M. Krause, Anton Josef Heckens and Yuriy Stepanov for helpful discussions.

\section*{Author contributions}

T.G. and M.S. proposed the research. S.W. and T.G. developed the methods of analysis. S.W. performed all the calculations. S.G. prepared the traffic data. All authors contributed equally to analyzing the results, writing and reviewing the paper.


\end{document}